\documentclass[namedreferences]{SolarPhysics}
\usepackage[optionalrh]{spr-sola-addons} % For Solar Physics

\usepackage{graphicx}        % For eps figures, newer & more powerfull
\usepackage{color}           % For color text: \color command
\usepackage{url}             % For breaking URLs easily trough lines
\renewcommand{\vec}[1]{ {\mathbf #1} }

\newcommand{\adv}{    {\it Adv. Space Res.}}

\newcommand{\aap}{    {\it Astron. Astrophys.}}

\newcommand{\aj}{     {\it Astron. J.}}
\newcommand{\apj}{    {\it Astrophys. J.}}

\newcommand{\jgr}{    {\it J. Geophys. Res.}}

\newcommand{\pasp}{   {\it Pub. Astron. Soc. Pac.}}

\newcommand{\solphys}{{\it Solar Phys.}}

\begin{document}

\begin{article}

\begin{opening}

\title{Characteristic features of the solar corona during the eclipse of 1 August 2008\\ {\it Solar Physics}}

\author{V.I.~\surname{Skomorovsky}\sep
        V.D.~\surname{Trifonov}\sep
        G.P.~\surname{Mashnich}\sep
        Yu.S.~\surname{Zagaynova}\sep
        V.G.~\surname{Fainshtein}$^{1}$\sep
        G.I.~\surname{Kushtal}\sep
        S.A.~\surname{Chuprakov}\sep
       }
\runningauthor{Skomorovsky et al.}
\runningtitle{K-corona in eclipse of 1 August 2008}

   \institute{$^{1}$ Institute of Solar Terrestrial Physics Russian Academy of Sciences, Siberian Branch,
                     email: \url{vfain@iszf.irk.ru}\\
             }

\begin{abstract}
In order to study the solar corona during eclipses, a new telescope was constructed. Three coronal images were obtained simultaneously from one objective of the telescope as the coronal radiation passed through three polarisers (whose transmission directions were turned through 0$^\circ$, 60$^\circ$, and 120$^\circ$ to the chosen direction); one image without polariser was also obtained. The telescope was used to observe the solar corona during the eclipse of 1 August 2008. We obtained distributions of the polarisation brightness, K-corona brightness, degree of the K-corona polarisation and total polarisation degree; polarisation direction depending on the latitude and radius in the plane of the sky was also obtained. We calculated radial distributions of electron density, depending on the latitude. Properties of all these distributions in different coronal structures were compared. We determined temperature of coronal plasma in different coronal structures on the assumption that there is a hydrostatic equilibrium.
\end{abstract}
\keywords{Corona, eclipse; K- and F- corona brightness; Polarization degree; Eclipse Observations}
\end{opening}
%-------------------------------------------------

\section{Introduction}
     \label{S-Introduction}

Observations of the solar corona have been regularly made over the last decades, both by ground-based and space coronagraphs and instruments designed to register the lower corona radiation in different spectral ranges (soft X-rays, extreme ultraviolet, etc.); coronal observations during eclipses remain, however, an effective tool for studying the solar corona.

  %{\S}{\bf --- Polarization of K- and F- corona} \\
One of the tasks being resolved due to observations of the solar corona during solar eclipses is to study polarisation features of the coronal radiation. The coronal glow in the continuous spectrum (the white-light corona) is known to be partially polarised. Up to $R\approx5R_{\rm o}$ ($\vec{R}_{\rm o}$ is the solar radius), polarisation of the white-light corona is dependent on the K-corona polarisation, since the F-corona is not polarised at these distances \cite{Koutchmy85}. Recall that the total surface brightness of the white-light corona ($\vec{B}$) is the sum of the K-corona brightness ($\vec{B}_{\rm K}$) and the F-corona brightness ($\vec{B}_{\rm F}$):

\begin{equation}\label{Eq-1}
B=B_{\rm K}+B_{\rm F}
\end{equation}

The K-corona forms from photospheric radiation scattered on free electrons of the corona; the F-corona, on particles of solar dust. Degree of the K-corona polarisation, being other than $1.0$, is determined by finite sizes of the Sun comparable to scales of property changes of the lower and middle corona and by distribution peculiarity of free-electron density along the line of sight.

  %{\S}{\bf --- Basic principle of polarizatiob telescopes, I method of coronal observations} \\
To study polarisation features of the white-light corona, a telescope is usually used; its entrance is equipped with a rotating polarisator (see, for instance, \cite{Kishonkov75}). In this case, coronal images recorded at various transmission directions of the polariser are in the same place of the recorder. In other words, coronal images obtained at various transmission directions of the polariser are recorded at different moments of time.

  %{\S}{\bf --- II method of coronal observations} \\
There is another technique to record polarised radiation: three coronal images are obtained simultaneously as the coronal radiation passes through three polarisers whose transmission directions are turned through 0$^\circ$, 60$^\circ$ and 120$^\circ$. Such a technique was employed by Yu.N. Lipsky et al. \cite{Shclovsky62} to observe the eclipse of 15 February 1961. A three-edged pyramid placed in front of the camera objective provided three coronal images in polarised beams at one exposure. When observing the same eclipse, N.S. Shilova \cite{Shilova61} used three objectives to obtain three coronal images. In front of the film emulsion that recorded radiation, there were three film polaroids with axes inclined at 120$^\circ$ to each other.

  %{\S}{\bf --- Advantage of the method} \\
Advantage of this technique when making polarimetric measurements in the corona is obvious: it allows us to record coronal images simultaneously (and not consecutively) at various positions of the transmission direction of polariser.

  %{\S}{\bf --- Our method and list of results in a few words} \\
This paper presents findings of investigation into the white-light corona during the solar eclipse of 1 August 2008. The findings were made with the telescope, using the analogous technique to record the coronal radiation and the original construction of the telescope. At this telescope, three coronal images were obtained simultaneously from one objective as the coronal radiation passed through three polarisers; one image without polariser was also obtained. We obtained polarisation brightness, K-corona brightness, polarisation degree, electron density, and coronal plasma temperature, depending on coordinates in the plane of the sky.

\section{Experimental part} %%%%%%%%%%%%%%%%%%%%%%%%%%%%%%%%%%%%%%%%
      \label{S-general}

\subsection{Telescope} %%%%%%%%%%%%%%
  \label{S-text}
The telescope scheme is presented in Fig. 1. A holder with four windows is placed in front of the telescope objective ($D=100$ mm, $F=250$ mm). Three windows contain polarisers; the forth window, a glass plate to smooth the reflection coefficient. Achromatic prisms that decline images to four matrix sectors are placed in each window. The prism consists of two wedges made of $K8$ and $F1$ glass. Chromatic aberration does not exceed the telescope's spatial resolution and is $\le 2.8^{\prime\prime}$. To remove the image overlay, we use a diaphragm (two thin foils perpendicular to each other, with a black material on them). Field of the telescope is $2.5^\circ\times3.5^\circ$. The 12-bit CCD matrix Hamamatsu C9300-124, with the chip of $24\times36$ mm ($2600\times4000$ pixels) was fixed in the focus.

\begin{figure}    %%%%%%%%%%%%%%%%%% FIGURE 1
   \centerline{\includegraphics[width=1.0\textwidth,clip=]{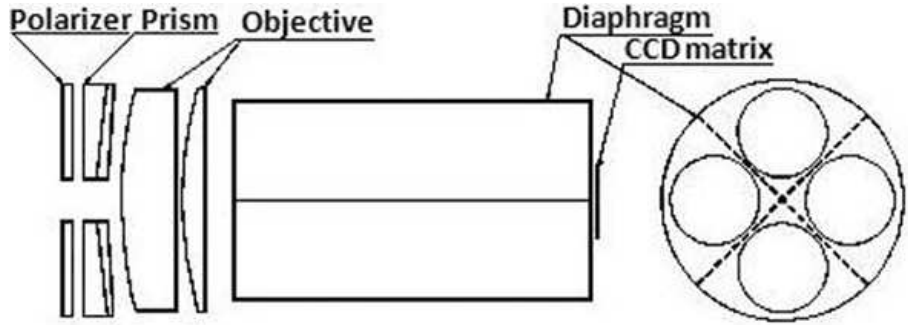}}
   \caption{The telescope scheme.}
   \label{F-simple}
\end{figure}

\subsection{Calibration} %%%%%%%%%%%%%%
  \label{S-labels}

We used three approaches to the telescope calibration. Two of them can be considered as a preliminary, less accurate calibration. They were used to calibrate the polarisation brightness. The third approach provided more accurate values of the measured polarisation brightness (\textbf{Pb}) and brightness of K-corona ($\vec{B}_{\rm K}$). The first approach is instrumental. It implied determining the filter transmission (used to measure the brightness of the Sun) as well as taking into account its spectral characteristics and other factors. The second approach comprised comparison of the measured polarisation brightness of the coronal radiation with calibrated measurements of the polarisation brightness in MLSO MarkIV coronagraph \url{(http://mlso.hao.ucar.edu/cgi-bin/mlso_homepage.cgi)} at the moment shifted by several hours about the eclipse time. Difference in results of the telescope calibration by the two methods was about 25\%. Below are results obtained with the use of the second approach. The third approach to the telescope calibration comprised reduction of the total measurable coronal brightness at the pole (at $R=1.7R_{\rm o}$) and equator (at $R=2R_{\rm o}$) to values of the total brightness of the white-light corona from the Koutchmy-Lamy model \cite{Koutchmy85} at these distances. Application of this approach was accompanied by correction of the polarisation brightness values obtained with the second approach to the calibration. This approach to the telescope calibration will be described in more detail in the following sections.

\subsection{Eclipse} %%%%%%%%%%%%%%
  \label{S-labels}

Observation of the eclipse of 1 August 2008 was made several tens of kilometres from Academy Town in Novosibirsk, on the shore of the Ob's sea. The observation place was near the central line of the eclipse band. Coordinates of the centre of Novosibirsk are $55^\circ02^\prime$ North, $82^\circ55^\prime$ East. Solar elevation at the moment of the eclipse was $30^\circ28^\prime$. According to calculations, the maximum phase of the eclipse was at $10:45$ UT. Duration of the total phase of the eclipse was about 2 minutes 23 seconds.

\subsection{Image preprocessing} %%%%%%%%%%%%%%
  \label{S-labels}

  %{\S}{\bf --- Data} \\
When making observations of the solar corona during the total solar eclipse of 1 August 2008, we obtained 274 images of the format *.TIFF. Among them, 194 images were with the exposure of 5.25 msec; 2 frames, with that of 7.0 msec; exposure of the last 78 images was 10 msec. Each image consists of four segments: 3 polarisation images and 1 image of the total coronal brightness without polariser. Availability of 4 coronal images on one frame allows us to study the corona up to $\approx3R_{\rm o}$. According to the data analysis, non-distorted images extend only to $\approx(1.7\div2)R_{\rm o}$.

  %{\S}{\bf --- Processing program} \\
Processing of data on the eclipse was made in IDL 6.1. Observations of Jupiter were used to debug the image overlay software.

  %{\S}{\bf --- Image preprocessing procedures} \\
Data processing package includes the following procedures: \\
  1) Elimination of the dark current obtained separately before the eclipse observations.\\
  2) Separation of images into 4 segments containing coronal images at different directions of the polariser axis.\\
  3) Correction for the flat field. Before beginning of the eclipse, we placed a special filter at the telescope entrance and made a survey of the sky at the zenith, with the exposures that were then used for making coronal observations during the eclipse. The coronal images were divided by the sky image averaged over 7 frames.\\
  4) Rough overlay of coronal images in the centre of the moon's disk. At this stage, the telescope resolution was defined more accurately ($\sim7^{\prime\prime}$).\\
  5) Image overlay with the use of fine structure elements (with polar plumes) to compensate motion of the moon across the solar disk, and value determination of small shifts of two images adjacent in time.\\
  6) Reduction of images in each polarisation to one exposure of 10.0 ms and overlay of images obtained in such a way. In this case, we took account of nonlinear variation in sensitivity of the recorder's elements in the domain of underexposures and overexposures at each frame.\\
  7) Matching of three channels to polarisers. As a result, 4 images were obtained: three coronal images ($image_{0^\circ}$, $image_{60^\circ}$ and $image_{120^\circ}$) and one coronal image for the window without polariser.

  %{\S}{\bf --- Using one image with an exposure of 10 ms} \\
Some coronal features turned out to be determined more accurately when one image is analysed at each position of the polariser with an exposure of 10 ms. For these images, we performed all necessary procedures of the image preprocessing, which were described above. Though these procedures were always performed carefully, the same measured parameters were different when we used all images or just one image for analysis. Values of measured parameters, averaged over data with the use of all coronal images obtained with different exposures and two separate images with an exposure of 10 ms, are used as results in the paper. Some results were obtained using one image with an exposure of 10 ms.

\section{Analysis results of the white-light corona images} %%%%%%%%%%%%%%%%%%%%%%%%%%%%%%%%%%%%%%%%
      \label{S-features}

\subsection{Coronal structure} %%%%%%%%%%%%%%
  \label{S-equations}
Investigation into the coronal structure at different periods of solar activity has been one of the main lines of research into the corona during solar eclipses for a long time \cite{Vsehvjatsky65,Loucif89}. According to the present notion, geometry of coronal structures is determined by the magnetic field. Nowadays three large-scale structures different in configuration of the magnetic field are distinguished. These are coronal holes (coronal regions with open field lines), coronal streamer belt (separating coronal holes with opposite polarity of the magnetic field), and chains of streamers (separating holes with the same field polarity). Bases of the belt and chains of streamers are formed by closed field lines with different configurations (see the review and monographs \cite{Zirker77,Schwenn90,Aschwanden04} as well as other original papers \cite{Hoeksema84,Eselevich99,Wang07}).

\begin{figure}    %%%%%%%%%%%%%%%%%% FIGURE 2
   \centerline{\includegraphics[width=1.0\textwidth,clip=]{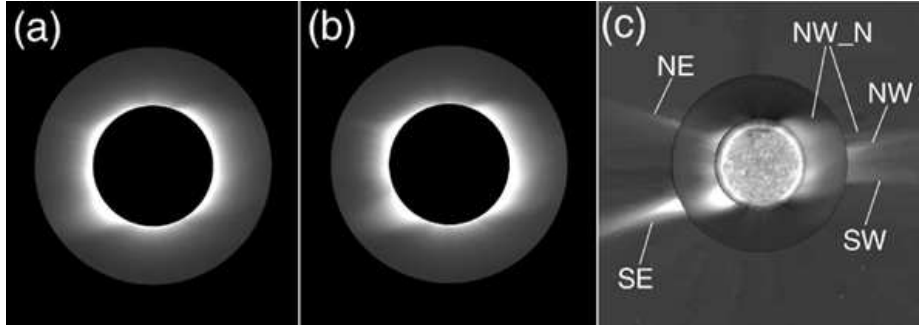}}
   \caption{(a) is the total coronal brightness; (b) - the polarisation brightness; (c) - the overlay of solar and coronal images (image of the Sun in the FeXII 195A extreme ultraviolet line (SOHO/EIT)), total brightness of the white-light corona (obtained with our telescope) and brightness of the white-light corona (according to the SOHO/LASCO C2 data). Figure presents non-calibrated images.}
   \label{F-simple}
\end{figure}

  %{\S}{\bf --- Total coronal brightness B in EIT/LASCO C2/Our images} \\
Using images during the eclipse of 1 August 2008, we tried to find out what coronal structures determine its glow on eclipse day. Fig. 2(a) presents non-calibrated distribution of the total coronal brightness $\vec{B}$ in the plane of the sky. Fig. 2(b) shows non-calibrated distribution of the brightness of polarised component of the white-light corona (polarisation brightness) $\vec{B}_{\rm P}$. Fig. 2(c) exhibits overlay of three images of the Sun and corona (image of the Sun in the $Fe~XII~195\AA$ extreme ultraviolet line (SOHO/EIT)), total brightness of the white-light corona (obtained with our telescope) and non-calibrated brightness of the white-light corona (according to the SOHO/LASCO C2 data). The total brightness of the white-light corona $\vec{B}$ and intensity of polarised radiation $\vec{B}_{\rm P}$ are expressed with formulae \cite{Billings66} in terms of intensities $B_{0^\circ}$, $B_{60^\circ}$ and $B_{120^\circ}$ recorded by three channels at different positions of the transmission direction of polariser:

\begin{equation}\label{Eq-21}
B = (2/3)(B_{0^\circ}+B_{60^\circ}+B_{120^\circ}),
\end{equation}
\begin{equation}\label{Eq-22}
B_P = (4/3)[(B_{0^\circ}-B_{60^\circ})B_{0^\circ}+(B_{60^\circ}-B_{120^\circ})B_{60^\circ}+(B_{120^\circ}-B_{0^\circ})B_{120^\circ}]^{1/2}.
\end{equation}

\begin{figure}    %%%%%%%%%%%%%%%%%% FIGURE 3
   \centerline{\includegraphics[width=1.0\textwidth,clip=]{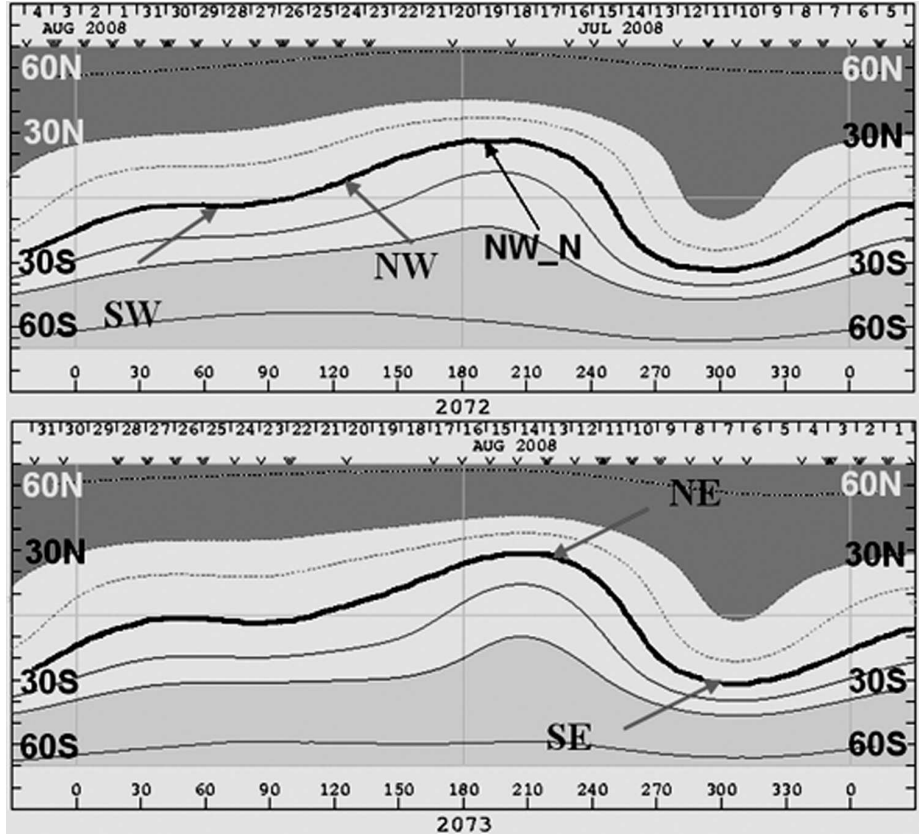}}
   \caption{Isolines of the magnetic field, including the neutral line (heavy line) on the source's surface ($R=2.5R_{\rm o}$), according to magnetic field calculations in the corona at WSO. Symbols $NE$, $SE$, $NW$, $SW$ and $NW_{-}N$ denote regions of the streamer belt, recorded as bright streamers in images of the white-light corona on Fig. 2(c).}
   \label{F-simple}
\end{figure}

  %{\S}{\bf --- Coronal structure specification at 1 August 2008} \\
It is seen from Fig. 2 that there are two polar coronal holes and several streamers during the solar eclipse. These streamers are the projections of different regions of the coronal streamer belt onto the plane of the sky. In order to determine these regions, let us consider the magnetic field neutral line (\textbf{NL}) on the source's surface ($R=2.5R_{\rm o}$), separating regions of the magnetic field with opposite polarity in the corona (Fig. 3). The \textbf{NL} configuration is obtained from magnetic field calculations in the corona in the potential approximation at WSO \url{(http://wso.stanford.edu/synsourcel.html)}. Shape of \textbf{NL} in the first approximation is known to be like that of the streamer belt \cite{Eselevich99,Wilcox83}. Fig. 2 and 3 present streamers and their corresponding \textbf{NL} regions (streamer belts), marked off by the same symbols (\textbf{SE}, \textbf{NE}, etc.). The brightest streamer \textbf{SE} (the south-eastern part of the corona) results from the projection of a part of the streamer belt (quasi-perpendicular to the plane of the sky, with its middle near the plane) onto the plane of the sky. Less bright streamers result from the projection of parts \textbf{NE}, (\textbf{SW} and $\vec{N}\vec{W}_{-}\vec{N}$ of the streamer belt also quasi-perpendicular to the plane of the sky, with their centres at an angular distance (along longitude) of $50^\circ\div60^\circ$ from the plane of the sky) onto the plane of the sky. A large region of increased brightness (or a group of streamers) (\textbf{NW} results from the projection of an extended region of the streamer belt NW onto the plane of the sky.  Noteworthy is the fact that all the streamers under consideration are non-radial. This is apparent from the coronal image during the eclipse, obtained with special method for isolating fine structure (see Fig. 1 in \cite{Habbal10}).

  %{\S}{\bf --- Saddle in coronal structure} \\
One more structural element of the corona in the latitudes in the vicinity of $\theta=20SE$ is also noteworthy, $\theta$ is the latitude counted from the equator. The term "saddle" is used in reference to it. This element of the coronal structure is the most pronounced in distribution of the total polarisation degree in the plane of the sky (Fig. 4А):

\begin{equation}\label{Eq-3}
P_{\rm total}(R,\theta)=Pb/B
\end{equation}

\begin{figure}    %%%%%%%%%%%%%%%%%% FIGURE 4
   \centerline{\includegraphics[width=1.0\textwidth,clip=]{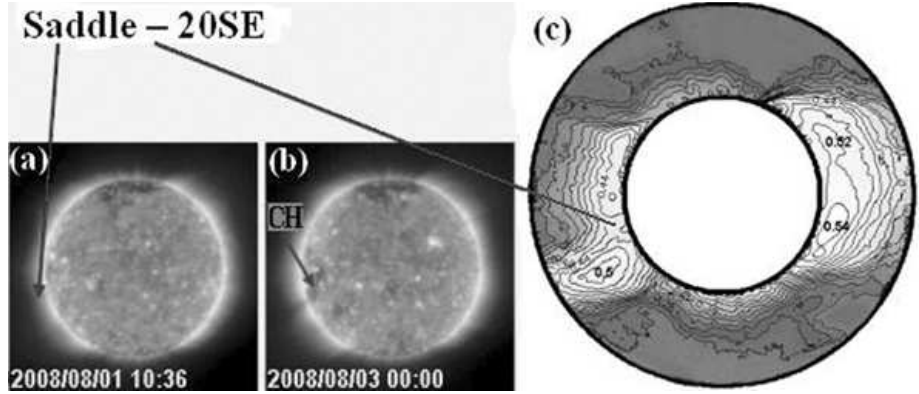}}
   \caption{(a,b) - solar images in the $195\AA$ line close to the moment of the eclipse (a) and 2 days after it (b), according to the SOHO/EIT data. Arrows indicate the saddle and its related coronal hole (CH); (c) is the distribution of the total polarisation degree of the coronal radiation $P_{total}=Pb(R,\theta)/B(R,\theta)$ in the plane of the sky}
   \label{F-simple}
\end{figure}

Here, non-calibrated \textbf{Pb} is the polarisation brightness, $\vec{B}$ is the total brightness of the white-light corona, calculated with formulae \cite{Billings66} in terms of intensities $B_{0^\circ}$, $B_{60^\circ}$ and $B_{120^\circ}$ recorded by three channels at different positions of the transmission direction of polariser. Though such a structure had been repeatedly recorded before, it turned out to be insufficiently studied. This structure was found out to occur between two streamers before appearance of a coronal hole on the visible solar disk. This is evident from Fig. 4(a,b) showing solar images in the  $Fe~XII~195\AA$ line (SOHO/EIT).

\subsection{Polarisation brightness as an intensity of polarized radiation} %%%%%%%%%%%%%%
  \label{S-equations}
To analyse intensity of polarised radiation, $B_{\rm P}(R,\theta)$ radial distributions were set up. Here, $\vec{R}$ is the radius in the plane of the sky, connecting the centre of the solar disk and the observation point. A preliminary calibration of $B_{\rm P}(R,\theta)$ distributions was made through their comparison with radial distributions of polarisation brightness $Pb_{\rm MkIV}(R,\theta)$ obtained at the Mark IV coronagraph (MLSO). Recall that the polarisation brightness $Pb={B_{\rm t}}-{B_{\rm r}}$, where $B_{\rm t}$ and $B_{\rm r}$ are the radiation intensities of the white-light corona polarised in the tangential and radial directions, respectively. Having regard to the fact that the F-corona is not polarised up to $R\approx5R_{\rm o}$, we have: $Pb={B_{\rm Kt}}-{B_{\rm Kr}}$, where $B_K$ is the K-corona brightness. It can be shown that $Pb=B_{\rm P}$ when the polarisation direction coincides with the tangential direction.

  %{\S}{\bf --- Data calibration using MarkIV} \\
Notice that the first measurements of $Pb_{\rm MkIV}(R,\theta)$ on 1 August 2008 started approximately 6 hours after the maximum phase of the solar eclipse. Nevertheless, comparison between $B_{\rm P}(R,\theta)$ and $Pb_{\rm MkIV}(R,\theta)$ is quite correct, since the corona changes insignificantly within 6 hours, provided that there are no strong transient disturbances (CME, etc.). Comparison of $Pb_{\rm MkIV}(R,\theta)$ radial distributions in several latitudes during 4 hours (the period when MarkIV was operating on 1 August 2008) shows that discrepancy between $Pb_{\rm MkIV}(R,\theta)$ distributions in all latitudes at different moments of time does not exceed $\sim15\%$ on average, and it depends mainly on noises. We used the following relation to calibrate our measurements:

\begin{equation}\label{Eq-4}
B_{\rm PC}(R,\theta)=B_{\rm P}(R,\theta)Pb_{\rm MkIV}(1.2R_{\rm o},\theta)/B_{\rm P}(1.2R_{\rm o},\theta).
\end{equation}

Here, $B_{\rm PC}(R,\theta)$ is the calibrated intensity of the polarised radiation, parameters $Pb_{\rm MkIV}(R,\theta)$ and $B_{\rm P}(1.2R_{\rm o},\theta)$ are the polarisation brightness (according to MkIV data) and intensity of polarised radiation at $R=1.2R_{\rm o}$, respectively.

  %{\S}{\bf --- Shorthand notation of polarisation brightness} \\
In what follows, we will use \textbf{Pb} to denote intensity of the polarised radiation of the white-light corona.

\begin{figure}    %%%%%%%%%%%%%%%%%% FIGURE 5
   \centerline{\includegraphics[width=1.0\textwidth,clip=]{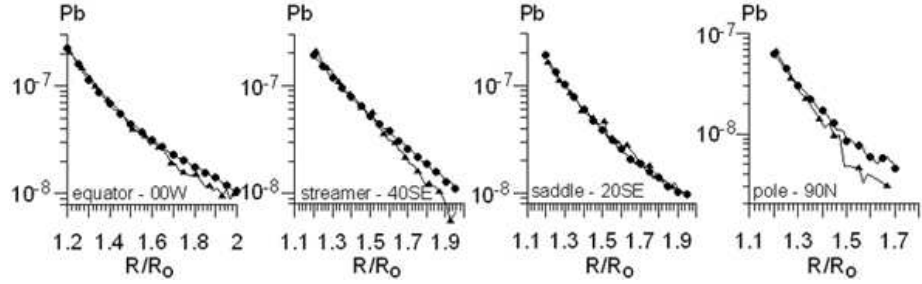}}
   \caption{Radial distributions of the polarisation brightness $Pb(R,\theta)$ obtained at the MarkIV coronagraph (triangle) and calibrated (preliminary calibration) distributions of the polarisation brightness obtained with the use of the new telescope (circle). MarkIV data are smoothed over high-frequency radial perturbations.}
   \label{F-simple}
\end{figure}

  %{\S}{\bf --- Predetermined data calibration} \\
To make comparison in several latitudes, we present (Fig. 5) $Pb(R,\theta)$ radial distributions obtained at the MarkIV coronagraph and polarisation brightness distributions obtained during the eclipse and calibrated with the said method. This calibration is called "preliminary", since an additional correction is required to match characteristics depending on \textbf{Pb} or influencing \textbf{Pb} (see section 3.4.).

\subsection{The K-corona brightness and total coronal brightness} %%%%%%%%%%%%%%
  \label{S-equations}
To find the K-corona brightness $B_{\rm K}(R,\theta)$ and the total coronal brightness $B_{\rm KF}=B_{\rm K}+B_{\rm F}$, we used another calibration, after having modified the method proposed by O.G. Badalyan in \cite{Badaljan88}. When applying this method, we took account of the fact that $B_{\rm F}>>B_{\rm K}$ in the polar region at $R>2R_{\rm o}$; thus, the total coronal brightness is $B_{\rm KF}\approx B_{\rm F}$. Using the comparison between the coronal brightness $B(R)$ observed in the polar region at $R\ge2R_{\rm o}$ with the F-corona brightness (according to the chosen model of this coronal feature), we deduced value $\bigtriangleup=lgB_F-lgB$. All brightness values obtained for the concrete eclipse were then corrected for this value. In this paper, we have chosen the F-corona model by Koutchmy-Lamy \cite{Koutchmy85} as a standard to obtain absolute intensities.

  %{\S}{\bf --- Calibration of observation brightness} \\
Unfortunately, we can not apply the method proposed in \cite{Shclovsky62} in this case, because coronal brightness measurements $B(R)$ in the polar region are reliable only up to $R\approx1.7R_{\rm o}$. If we use values $\bigtriangleup=lgB_{\rm F}-lgB$ (deduced at  $R\ge2R_{\rm o}$) to correct the brightness under observation, we will then deduce the K-corona brightness $B_K$ with large errors. We therefore modified the method from \cite{Badaljan88} in the following way.  First of all, we divided the coronal region under observation into two parts: the polar region ($\mid\theta\mid>60^\circ$) and the near-equatorial region ($\mid\theta\mid>60^\circ$). Recall that $\theta$ is the latitude counted from the equatorial plane; $\theta>0$ with distance away from the equator to the centre, $\theta<0$ to the south of the equator. In the polar region, we calibrated the brightness under observation in the following manner:

\begin{equation}\label{Eq-5}
B(R,\theta)_{\rm calib}=[(B_{\rm K}(1.7R_{\rm o})+B_{\rm F}(1.7R_{\rm o}))_{\rm KL-POLE}/B(1.7R_{\rm o}, \theta=90^\circ)]B(R,\theta).
\end{equation}

Here, $(B_{\rm K}(1.7R_{\rm o})+B_{\rm F}(1.7R_{\rm o}))_{\rm KL-POLE}$ is the total coronal brightness in the Koutchmy-Lamy model \cite{Koutchmy85} at the pole, at $R=1.7R_{\rm o}$. Calibration in the near-equatorial region is as follows:

\begin{equation}\label{Eq-6}
B(R,\theta)_{\rm calib}= [(B_{\rm K}(2R_{\rm o})+B_{\rm F}(2R_{\rm o}))_{\rm KL-EQUATOR}/B(2R_{\rm o},\theta=0^\circ)]B(R,\theta).
\end{equation}

We chose $B(2R_{\rm o},\theta=0^\circ)$ in the western part of the corona for calibration. Expression $(B_{\rm K}(2R_{\rm o})+B_{\rm F}(2R_{\rm o}))_{\rm KL-EQUATOR}$ is the total coronal brightness in the Koutchmy-Lamy \cite{Koutchmy85} model at the equator.

  %{\S}{\bf --- K-corona brightness} \\
Having deduced calibrated values of the total coronal brightness $B(R,\theta)_{\rm calib}$, we determined the K-corona brightness $B_{\rm K}$ with the use of the relation:

\begin{equation}\label{Eq-7}
B_{\rm K}(R,\theta)=B(R,\theta)_{\rm calib}-B_{\rm F-KL}(R).
\end{equation}

Here, $B_{\rm F-KL}(R)$ is the F-corona brightness from the Koutchmy-Lamy model \cite{Koutchmy85}. At $R\le2R_{\rm o}$, $B_{\rm F-KL}(R)$ radial distributions at the equator and pole are similar. When finding $B_{\rm К}(R,\theta)$ in both polar and near-equatorial regions, we used the $B_{\rm F-KL}(R)$ distribution average for the pole and equator.

  %{\S}{\bf --- Radial scans of brightnes of K-corona and total brightnes} \\
Fig. 6 presents radial distributions of the K-corona brightness (obtained by the said method) and of the total coronal brightness $B_{\rm KF}=B_{\rm K}+B_{\rm F}$ for some coronal regions. In this case, $B_{\rm KF}=B(R,\theta)_{calib}$. For comparison purposes, the figure also shows distributions of $B_{\rm K-KL}(R)$ and $B_{\rm KF-KL}=B_{\rm K-KL}(R)+B_{\rm F-KL}(R)$ at the pole and equator (taken from \cite{Koutchmy85}). It is apparent that the difference between distributions of $B_{\rm K}(R)$ and $B_{\rm KF}(R)$ at the equator and pole obtained for the eclipse of 1 August 2008  and model distributions of $B_{\rm K-KL}(R)$ and $B_{\rm KF-KL}(R)$ from \cite{Koutchmy85} is relatively small, whereas distributions of $B_{\rm KF}(R)$ and $B_{\rm KF-KL}(R)$ at the pole virtually coincide.

\begin{figure}    %%%%%%%%%%%%%%%%%% FIGURE 6
   \centerline{\includegraphics[width=1.0\textwidth,clip=]{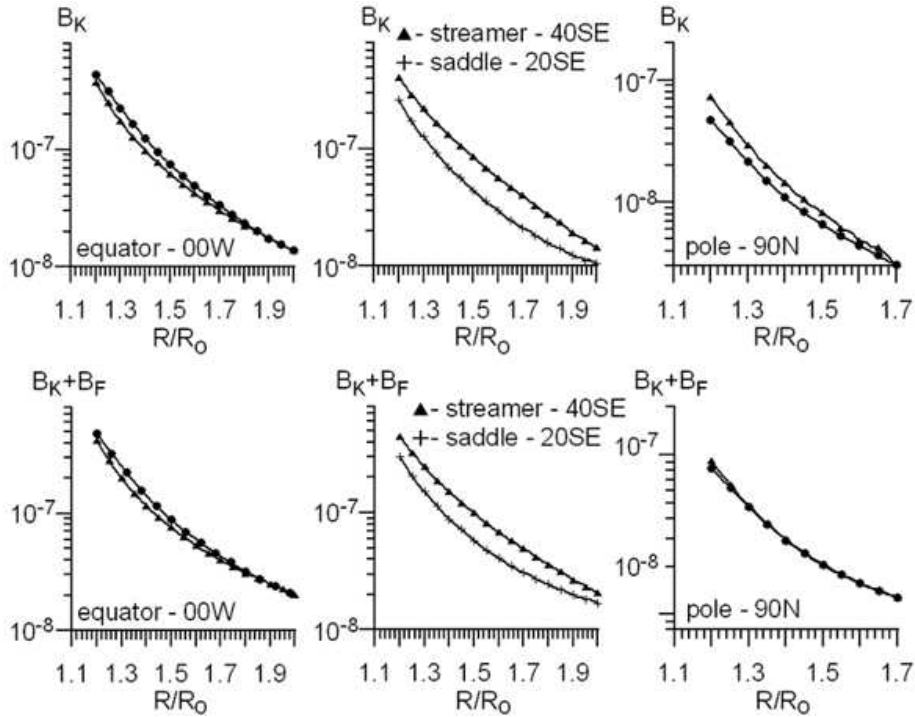}}
   \caption{Examples of radial distributions of the K-corona brightness $B_{\rm K}(R)$ and total coronal brightness $B_{\rm KF}(R)=B_{\rm K}(R)+B_{\rm F}(R)$ (solid lines with triangles). Lines with solid circles denote $B_{rm K-KL}(R)$ distributions at the pole and equator from (Koutchmy and Lamy, 1985). Lines with stars denote $B_{\rm K}(R)$ distributions obtained by the method based on calculations of the electron density.}
   \label{F-simple}
\end{figure}

  %{\S}{\bf --- Comments for method to determin of K-corona brightness} \\
Let us now comment on the above-described method for finding the K-corona brightness. If we deal with high-quality data, then, having reduced the brightness of the white-light corona measured at given distances at the pole and equator to values $B_{\rm K}+B_{\rm F}$ from the Koutchmy-Lamy model \cite{Koutchmy85}, we will deduce values $B_{\rm K}(R)$ equal to values of the K-corona brightness from \cite{Koutchmy85} at all distances. At the same time, the K-corona brightness in the same latitudes is known to vary significantly with time and to be different from the model described in \cite{Koutchmy85} in some cases. This implies that, using this computational technique for $B_{\rm K}(R)$, we suppose the error (unknown in the general case) at the pole and equator, used to find the K-corona brightness. This suggests that the additional correction of $B_{\rm K}(R)$ is required to bring this value into proximity with true values. A possible technique for making this correction will be discussed in the following section. Notice that values of different parameters of coronal plasma (electron density distributions, degree of the K-corona polarisation, plasma temperature) described below and obtained with the use of $B_{\rm K}(R)$ indicate satisfactory accuracy of our determination of the K-corona brightness. This is probably caused by the fact that the eclipse took place during the solar minimum, and the Koutchmy-Lamy \cite{Koutchmy85} is also concerned with the solar minimum.

  %{\S}{\bf --- One more method of determination of K-corona brightness} \\
We emphasize that one more method for determining the K-corona brightness is proposed in \cite{Saito77}. According to this method, concentration distribution $N_{\rm e}(R)$ is deduced from the $Pb(R)$ distribution in the given latitude; $N_{\rm e}(R)$ is then used to find $B_{\rm K}(R,\theta)$ through substitution of $N_{\rm e}(R)$ into the integral relation connecting $B_{\rm K}(R,\theta)$ to $N_{\rm e}(R)$ \cite{Billings66,Hundhausen93}. Advantage of this method is the absence of connection between the measured coronal brightness and the total brightness from the Koutchmy-Lamy model \cite{Koutchmy85} that contains both the F-corona brightness (varying slightly during the solar cycle) and the K-corona brightness (varying significantly with time at the equator and pole). We pretested this method (see methods applied to calculate $N_{\rm e}(R,\theta)$ and calculation results of electron density in the section 3.6.). Having made correction of $Pb(R,\theta)$ and of $B_{\rm K}(R,\theta)$ distributions necessary to obtain realistic values of degree of the K-corona polarisation (see the next section), we concluded that, through the application of this method, we obtain values of the K-corona brightness different from those obtained with the modified method \cite{Badaljan88} by the value within ($10\div30\%$).

\subsection{K-corona polarization degree and total degree of the white-light corona polarization} %%%%%%%%%%%%%%
  \label{S-equations}
Let us now analyse distributions of degree of the K-corona polarisation $P_{\rm K}=Pb/B_{\rm K}$ and of the total polarisation degree $P_{\rm KF}=Pb/B_{\rm KF}$ obtained with the use of calibrated values \textbf{Pb} and $\vec{B}_{\rm K}$. There are many papers dealing with research into polarisation degree of the coronal radiation. Here, we will mention only a few works \cite{Hulst50,Koutchmy77,Badalyan97,Koutchmy94,Badalyan08,Molodensky73,Nikolsky77,Duerst82,Kim96}, describing several eclipses. Notice that there are very few papers that present $P_{\rm K}(R)$ distributions deduced only from brightness measurements of the white-light corona (including intensities of the coronal radiation passing through polarisers), like in our case. Up to now, the paper by van de Hulst \cite{Hulst50} has been "canonical" paper of this type.

  %{\S}{\bf --- Anomalous big values of the K-corona polarization degree} \\
If we use $Pb(R)$ and $B_{\rm K}(R)$ values, deduced like those in Fig. 5 and 6, to set up $P_{\rm K}(R)$, we will obtain abnormally large maximum values of $\vec{P}_{\rm K}$ in many latitudes. We think that this results from inaccurate calibration of the polarisation brightness \textbf{Pb} that we obtained by comparing $Pb(R)$ values measured during the eclipse with values of the polarisation brightness deduced from MarkIV coronal observations on eclipse day. We have come to this conclusion due to the comparison of radial distributions of electron densities $N_{\rm e}(R)$,obtained with the use of $Pb(R)$ and $B_{\rm K}(R)$, in given latitudes. We have found out that distributions of Ne(R) obtained by the two methods as well as mean values of these distributions $<N_{\rm e}(R)>$ at $R=(1.2-(1.7\div2))R_{\rm o}$ differ noticeably.  We can calibrate the polarisation brightness more accurately using the relation:

\begin{equation}\label{Eq-8}
Pb(R)_{calib2}=\frac{<N_{\rm e}(R)>_{\rm BK}}{<N_{\rm e}(R)>_{\rm Pb}}Pb(R).
\end{equation}

Here, $<N_{\rm e}(R)>_{\rm BK}$ and $<N_{\rm e}(R)>_{rm Pb}$ are the values of electron density averaged over radius and obtained from the radial distribution of the K-corona brightness and polarisation brightness calibrated in accordance with MarkIV data. Having made this calibration of the polarisation brightness, we have:

\begin{equation}\label{Eq-8a}
<N_{\rm e}(R)>_{\rm BK}\approx <N_{\rm e}(R)>_{\rm Pb(R)_{calib2}}.
\end{equation}

  %{\S}{\bf --- Another approach to determine Ne(R)} \\
Nevertheless, we used another approach to correct $Pb(R)$. To compare correctly $P_{\rm K}(R)$ distributions we have deduced at the equator and pole with corresponding distributions of degree of the K-corona polarisation from \cite{Hulst50}, we reduced $\vec{P}_{\rm K}$ values (obtained with the use of Pb calibrated according to MarkIV data) at $R=1.2R_{\rm o}$ to values of $P_{\rm K}= P_{\rm KVDH}$ at these distances from van de Hulst \cite{Hulst50}. In the polar region, we used $P_{\rm K}(1.2R_{\rm o})= P_{\rm KVDH}(1.2R_{\rm o})$ obtained at the pole in \cite{Hulst50}; in the near-equatorial region, at the equator. Correction procedure (new calibration) of the polarisation brightness implied multiplying of $Pb(R)$ by coefficient

\begin{equation}\label{Eq-9a}
C=P_{\rm KVDH}(1.2R_{\rm o})/P_{\rm K}(1.2R_{\rm o}),
\end{equation}

therefore

\begin{equation}\label{Eq-9b}
Pb(R)_{calib}=C\cdot Pb(R).
\end{equation}

As a result of this procedure, we also obtain:

\begin{equation}\label{Eq-9c}
<N_{\rm e}(R)>_{\rm BK}\approx <N_{\rm e}(R)>_{\rm Pb_{calib}},
\end{equation}

and, in general, radial distributions of $N_{\rm e}(R)\cdot B_{\rm K}$ and $N_{\rm e}(R)\cdot Pb_{\rm calib}$ differ insignificantly! This fact will be described in Section 3.6.

  %{\S}{\bf --- Polarization degree in different structures of solar corona} \\
Fig. 7 shows $P_{\rm K}(R)$ distributions at the equator, streamer, saddle and pole. Figure presents 2 curves for both the equator and saddle.  The lower curves (depicted by lines with circles) are the averaging result of degree of the K-corona polarisation obtained from the entire set of coronal images during the eclipse with all exposures and 2 separate images with an exposure of 10 ms. The upper curves (that turned out to be the closest to the $P_{\rm K}(R)$ dependence at the equator from \cite{Hulst50} ($P_{\rm KVDH}$)) are obtained with the use of separate and different images with an exposure of 10 ms. This difference between curves $P_{\rm K}(R)$ obtained by different methods reflects the following fact. Degree of the K-corona polarisation (including its pattern of change with distance) is very sensitive to the determination accuracy of $Pb(R)$ and $B_{\rm K}(R)$. The spread of $P_{\rm K}(R)$ distributions in two types of coronal images implies that $Pb(R)$ and $B_{\rm K}(R)$ are found with the error influencing determination accuracy of $P_{\rm K}(R)$. The error is caused, to a certain degree, by the impossibility to obtain identical images due to the primary processing of different coronal images. This reflects complexity and, probably, insufficient accuracy of the methods of primary processing of images used by us.

\begin{figure}    %%%%%%%%%%%%%%%%%% FIGURE 7
   \centerline{\includegraphics[width=1.0\textwidth,clip=]{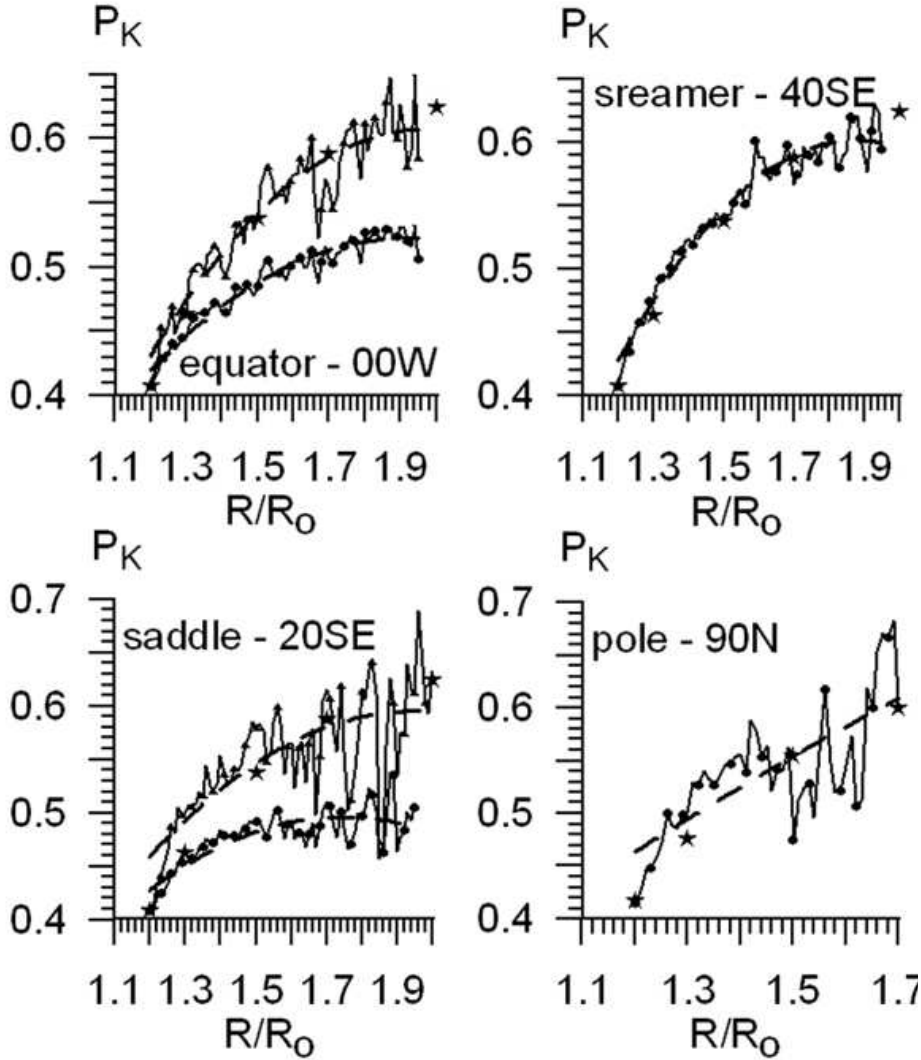}}
   \caption{Examples of radial distributions of degree of the K-corona polarisation $P_{\rm K}(R,\theta)$. Solid lines with circles denote our calculations of $\vec{P}_{\rm K}$ for the eclipse of 1 August 2008; the star, degree of the K-corona polarisation from (van de Hulst, 1950). Along with averaged $P_{\rm K}(R)$, distributions of degree of the K-corona polarisation (that are closest to the $P_{\rm K}(R)$ dependence from (van de Hulst, 1950) and obtained using one image with an exposure of 10 ms) are denoted by lines with triangles for the streamer and saddle, using several types of images.}
   \label{F-simple}
\end{figure}

  %{\S}{\bf --- Velue variation of K-corona polarization degree} \\
Having deduced $P_{\rm K}(R)$ distributions for the equator and saddle (that turned out to be close to the $P_{\rm KVDH}$ distribution at the equator), we do not  suppose that the distribution from \cite{Hulst50} should take place, for instance, at the equator at any moment of time. Actually, this is not true, and $P_{\rm K}(R)$ variations may take place (according to analysis). Different $P_{\rm K}(R)$ radial dependences (presented in the first approximation in figures on the left) bound regions within which true distributions of degree of the K-corona polarisation are found.

  %{\S}{\bf --- Variation of total polarization degree} \\
Fig. 8 presents distributions of the total polarisation degree $P_{\rm KF}(R)$ at the equator, streamer, saddle and pole, and $P_{\rm KF}(R)$ distributions obtained in the streamer and polar region during the eclipse of 30 June 1973 \cite{Koutchmy77,Badalyan97} are shown for comparison. Notice the clear difference (especially at the pole) between total polarisation degrees obtained by 2 research teams when observing the same eclipse of 30 June 1973. This is one more circumstantial evidence of the fact that acquisition of reliable values of the total polarisation degree and especially of degree of the K-corona polarisation is a very difficult task.

\begin{figure}    %%%%%%%%%%%%%%%%%% FIGURE 8
   \centerline{\includegraphics[width=1.0\textwidth,clip=]{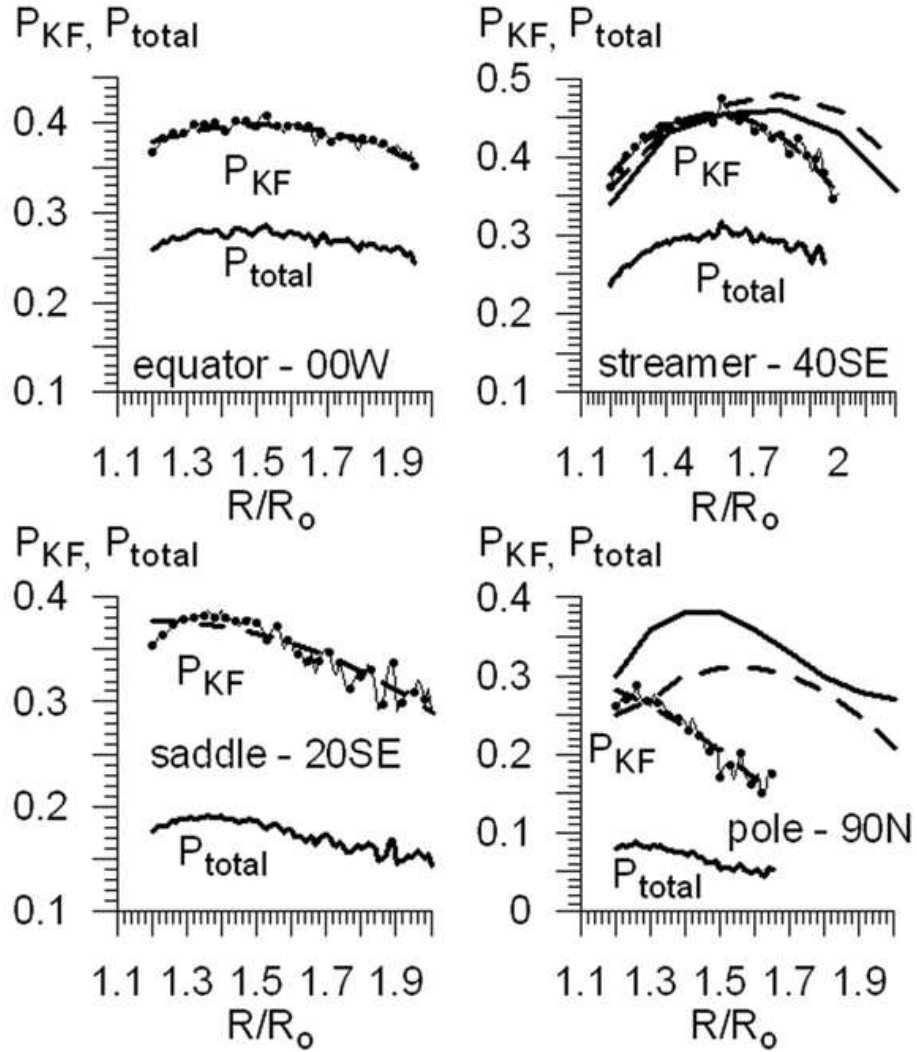}}
   \caption{Examples of radial distributions of the total degree of the white-light corona polarisation $P_{\rm KF}(R,\theta)$ and $P_{\rm total}(R,\theta)$. Solid lines with circles denote our calculations of $P_{\rm KF}$ for the eclipse of 1 August 2008; heavy solid line, the total polarisation degree $P_{\rm total}$, taking into account the contribution of spurious radiation to the recorded signal. Solid and dotted lines show $P_{\rm KF}(R,\theta)$ distributions for the eclipse of 30 June 1973 from (Koutchmy, Picat, and Dantel,1977; Badalyan, Livshits, and Sycora, 1997).}
   \label{F-simple}
\end{figure}

  %{\S}{\bf --- Total polarization degree} \\
It is worth noting that the total polarisation degree in the literature is sometimes considered to mean value $P_{\rm total}=Pb/B$, where \textbf{B} is the total measurable brightness of the white-light corona comprising spurious radiation of a different nature \cite{Koutchmy94}. This value is thus physically insignificant. In spite of this fact, it turned out to be used to separate the K-corona brightness from the total brightness of the white-light corona under certain conditions \cite{Koutchmy94}. With this in mind, we give examples of $P_{\rm total}(R)$ for the eclipse of 1 August 2008 in Fig. 8. When finding these dependences, we used $Pb(R)$ values adjusted for coefficient $\vec{C}$.

\begin{figure}    %%%%%%%%%%%%%%%%%% FIGURE 9
   \centerline{\includegraphics[width=1.0\textwidth,clip=]{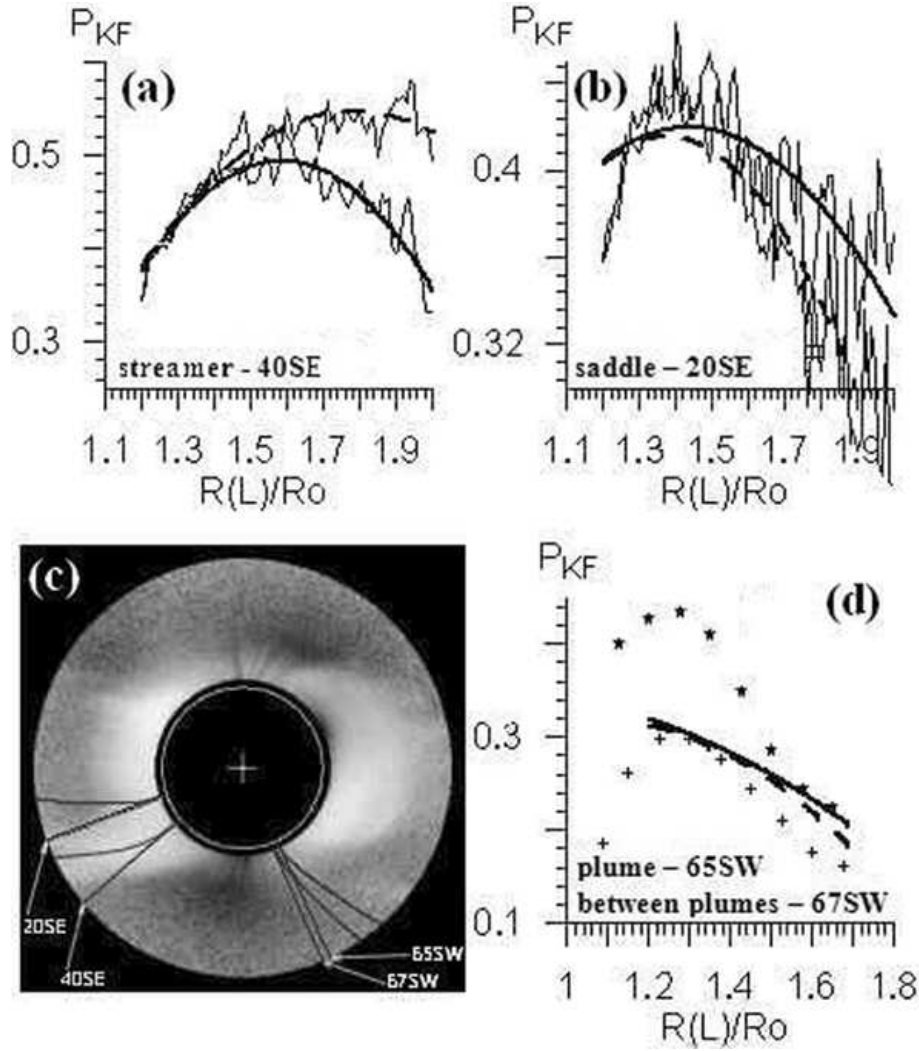}}
   \caption{(a,b) are the comparison of $P_{\rm KF}(L)$ distributions along the SE streamer and SE saddle (dotted lines, position of the point along the curved line is denoted by L) with radial (solid lines, position of the point along the trajectory is denoted by R) distributions of the total polarisation degree $P_{\rm KF}(L)$ in these structures, passing through centres of structures near the Sun; (c) - "trajectories" from which distributions of polarisation degree were deduced; (d) - comparison between our $P_{\rm KF}(L)$ distributions in the plume and between plumes, and in the plume (star) and between plumes (cross) for the eclipse of 3 November 1994 (Koutchmy, 1994). In order to avoid overload of the figure, the polarisation degrees obtained by us are presented only as second-order regression lines.}
   \label{F-simple}
\end{figure}

  %{\S}{\bf --- Radial scans and scans along line of symmetry of coronal structures} \\
All the streamers and the saddle we observed in the southeastern part of the corona (\textbf{20SE}) are nonradial (Fig. 4). Fig. 9 (a, b) illustrates distributions of the total polarisation degree $P_{KF}$ along the streamer (\textbf{SE}) and saddle (\textbf{SE}). Scan trajectories are presented in Fig. 9(c). These figures also show radial distributions of the polarisation degree in these structures, passing through centres of structures near the Sun. Fig. 9(d) presents distributions of the total polarisation degree $P_{\rm KF}(R)$ for the plume and space between plumes. In all these cases, polarisation degree was determined using one image with an exposure of 10 ms. For comparison purposes, the figure also shows $\rm P_{KF}(R)$ distributions in the plume and between plumes from \cite{Badalyan08}. In our case, difference in behaviour of the total polarisation degree in the plume and in adjacent region between plumes was insignificant, whereas that in \cite{Koutchmy94} was quite significant. We think that such a considerable difference does not reflect the real difference in $P_{\rm KF}(R)$ in plumes and between them, since the latter must be less significant. This conclusion can be substantiated due to comparison of distributions of electron density in plumes and between them. Such a detailed analysis is beyond the scope of this paper.

  %{\S}{\bf --- Fluctuation of polarization degree} \\
Let us note one peculiarity of Fig. 7-9. In these figures, $P_{\rm K}(R)$ and $P_{\rm KF}(R)$ distributions are characterised by high-amplitude fluctuations. Single images mainly contribute to these fluctuations. Even averaging of such images together with those involving all images with different exposures does not decrease the fluctuation level of the polarisation degree noticeably.

  %{\S}{\bf --- Why using labels ?} \\
In summary, let us propose a technique for additional correction of the K-corona brightness to obtain its more exact values, at least at the equator and pole.  Though $P_{\rm K}(R)$ in these coronal regions may differ from $P_{\rm KVDH}(R)$ distributions (in \cite{Hulst50}), there is a reason to believe that this difference is relatively small. In this case, $B_{\rm K}(R)$ may be corrected with the relation:

\begin{equation}\label{Eq-5}
B_{\rm K-corr}(R)=A\cdot B(R,\theta)_{\rm calib}-B_{\rm F-KL}(R).
\end{equation}

Here, \textbf{A} is the coefficient which is chosen under the condition that the $P_{\rm K}(R)$ distribution (being found with the use of $B_{\rm K-corr}(R)$) is the closest to distributions $P_{\rm KVDH}(R)$. Additional correction of $Pb(R)$ will also be required to match electron density distributions obtained from  $P_{\rm K}(R)$ and $Pb(R)$.

\subsection{Polarization direction} %%%%%%%%%%%%%%
  \label{S-equations}
To find polarisation direction, we used one coronal image obtained from three channels at different directions of the polariser axis with the exposure time of 10 ms. The polarisation direction was determined with the use of formula in \cite{Billings66}:

\begin{equation}\label{Eq-6}
\theta=(1/2)\arctan[3^{1/2}(B_{60^\circ}-B_{120^\circ})/(2B_{0^\circ}-B_{60^\circ}-B_{120^\circ})]+n\pi/2, n=0, \pm1, \pm2…
\end{equation}

  %{\S}{\bf --- What is the theta?} \\
Here, angle $\theta$ is counted from direction of the first polariser's axis turned through $0^\circ$ to the North pole. When choosing n, we took account of the fact that the polarisation direction differs insignificantly from the tangential direction.

  %{\S}{\bf --- Polarization direction isolines} \\
Fig. 10(e) presents isolines $\theta$. Fig. 10(a-d) shows changes in the polarisation direction dependent on the latitude for 4 quarters in latitude. In each latitude, these directions are averaged over radius in the range $R=(1.15\div1.3)R_{\rm o}$. The deviations of measured polarisation directions ($\vec{q}$) (that may be up to $10^\circ\div15^\circ$) from  "theoretical" values are apparently caused by insufficiently high quality of our coronal images. Actually, contribution of well-known physical causes leading to deviation of measured values q from "theoretical" values is within several degrees at $R\le2R_{\rm o}$ \cite{Molodensky73,Molodensky09}. Notice that there are methods for determining the polarisation direction of the coronal radiation with an accuracy to $1^\circ-2^\circ$ \cite{Kishonkov75}. Deviation of the measured polarisation directions from "theoretical" ones may be also caused by the fact that we did not take account of the additional polarisation resulting from the passage of radiation through the Earth's atmosphere.

\begin{figure}    %%%%%%%%%%%%%%%%%% FIGURE 10
   \centerline{\includegraphics[width=1.0\textwidth,clip=]{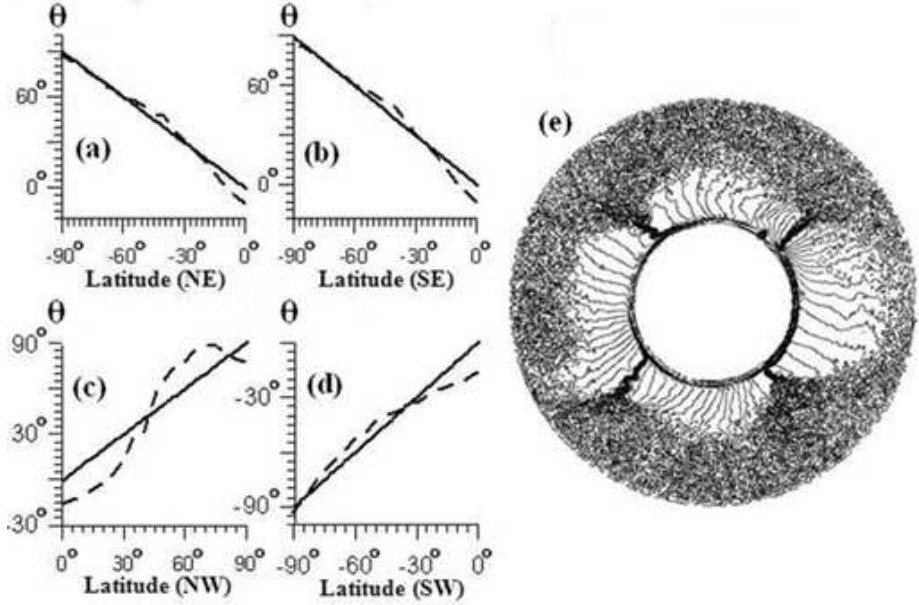}}
   \caption{Fig. 10.(a,b,c,d) are changes in the polarisation direction depending on the latitude for 4 quarters in latitude: $NE$, $SE$, $NW$, and $SW$. Solid line shows the tangential direction of polarisation; dotted, measurements carried out using one image with an exposure of 10 ms. (e) - contours of the polarisation direction image.}
   \label{F-simple}
\end{figure}

\subsection{Electron density in different coronal structures}%%%%%%%%%%%%%%
  \label{S-equations}
Let us first of all describe briefly the inversion method \cite{Hayes01} used to find electron density of coronal plasma $N_{\rm e}$. The method relies on the following assumptions: (1) plasma in the corona is spherically or axially symmetric; 2) electron density may be presented as a polynomial:

\begin{equation}\label{Eq-10}
N_{\rm e}(r)=\sum_{\rm k}\alpha_{\rm k} r^{\rm -k}, {\rm k}=1...n.
\end{equation}

Here, $\vec{r}$ is the radius connecting the centre of the solar disk and the observation point. Substituting this expression for $N_{\rm e}(r)$ into integral relations (that connect $Pb(R)$ or $B_{\rm K}(R)$ distributions with $N_{\rm e}(r)$ \cite{Billings66,Hundhausen93}), coefficients ak are selected in such a way as to minimise differences between measured $Pb(R,\theta)$ or $B_{\rm K}(R,\theta)$ distributions and calculated $Pb_{\rm calib}(R,\theta)$ or $B_{\rm K-calib}(R,\theta)$, using dependence $N_{\rm e}(r)=\sum_{\rm k}\alpha_{\rm k} r^{\rm -k}$.

    %{\S}{\bf --- Distribution Electron density in different coronal structures} \\
Using distributions of the polarisation brightness \textbf{Pb} and K-corona brightness $\vec{B}_{\rm K}$, we found radial distributions of the $N_{\rm e}(R,\theta)$ electron density in different latitudes, using method \cite{Hayes01}. Fig. 11 shows $N_{\rm e}(R,\theta)$ distributions in several coronal structures obtained in calculations with $n=6$ (we also performed calculations with $n=9$, but they provided reliable results only up to $R=(1.5\div 1.7)R_{\rm o})$.  We are going to compare electron densities in different latitudes by comparing Ne values at $R=1.2R_{\rm o}$ ($N_{\rm e}(R,\theta)$ distributions in Fig. 11 are presented only for some latitudes). The densest plasma is in the proximity of the equator up to the latitude $\approx(45^\circ\div 50^\circ)$. Electron density on the western limb is about $15\div20\%$ higher than that on the eastern limb. Variations of electron density in different coronal structures in the given latitude region (various streamers, saddle) are within $(10\div20)\%$. it should be noted that there is a decrease in the electron density compared to the equator and neighbouring streamer. When approaching the poles, the electron density becomes lower: it is about 4 times less at the pole than in the near-equatorial region.

\begin{figure}    %%%%%%%%%%%%%%%%%% FIGURE 11
   \centerline{\includegraphics[width=1.0\textwidth,clip=]{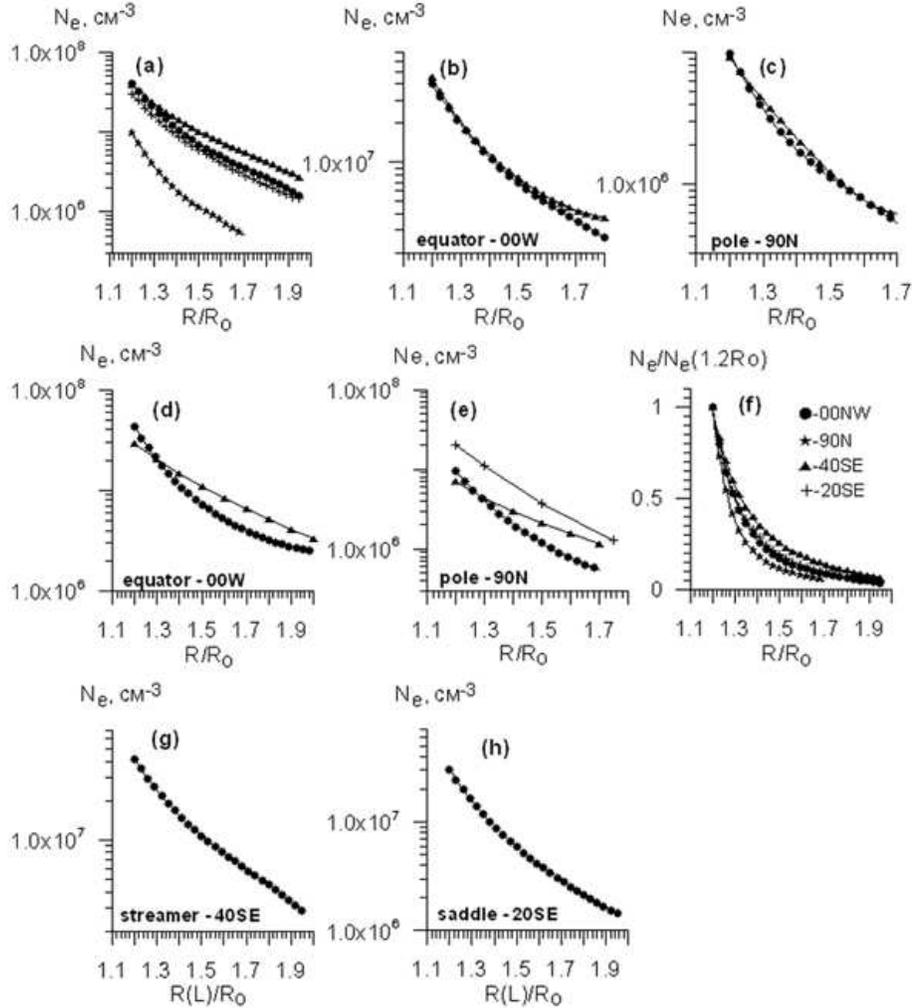}}
   \caption{(a) is radial distributions of electron density $N_{\rm e}(R)$ obtained from $B_{\rm K}$, at different latitudes. Dots, triangles, crosses, and stars correspond to latitudes $00W$, $20SE$, $40SE$, $90N$. (b,c) $-$ comparison of $N_{\rm e}(R)$ obtained from $B_{\rm K}$ and $Pb$. (d,e) $-$ comparison of radial distributions of electron density in the $SE$ streamer and at the pole (averaged over those obtained from $B_{\rm K}$ and $Pb$) $-$ dots, with $N_{\rm e}(R)$ distributions for other periods of time from (Hayes, Vourlidas, and Howard, 2001) (averaged over those obtained from $B_{\rm K}$ and $Pb$) $-$ triangles and from (Doyle, Teriaca, and Banerjee, 1999) $-$ crosses. (f) $-$ radial distributions of electron densities (obtained from $B_{\rm K}$) in different latitudes, normalised to $N_{\rm e}(1.2R_{\rm о})$. (g,h) $-$ comparison of $N_{\rm e}(R)$ radial distributions (using 1 image) and along structures($N_{\rm e}(L)$) in the streamer $40SE$ and saddle $20SE$.}
   \label{F-simple}
\end{figure}

  %{\S}{\bf --- Electron density at equator and pole} \\
Fig. 11(b,c) illustrates similarity between radial distributions of electron density obtained from distributions of the polarisation brightness and K-corona brightness at the equator and pole. Such a similarity is typical for other coronal regions too. Determination of electron density was made from distributions of the polarisation brightness, using \textbf{Pb} values calibrated twice (see Section 3.4.).

  %{\S}{\bf --- Electron density at equator and pole of another athors} \\
Fig. 11(d,e) demonstrates $N_{\rm e}(R,\theta)$ distributions obtained at the equator and pole and compared to distributions of electron density deduced for the same coronal regions in \cite{Hayes01,Doyle99}. Using data from \cite{Hayes01}, we averaged $N_{\rm e}(R)$ distributions over the polarisation brightness and K-corona brightness during the eclipse of 26 February 1998; the distributions were obtained by authors of this paper and corrected with the use of $N_{\rm e}(R)$ calculations according to the LASCO C2 coronagraph data at $R>2R_{\rm o}$. In \cite{Doyle99}, the $N_{\rm e}(R)$ distribution was obtained from UVCS/SOHO data.

  %{\S}{\bf --- Electron density at different structures of solar corona} \\
Fig. 11(f) shows $N_{\rm e}(R)$ relative distributions in different coronal structures.  An important physical feature of the corona is a rapid decrease in electron density with distance. This value is related to the representative plasma temperature, speed of the solar wind that forms in the corona, etc. According to expectations, the most rapid decrease in $N_{\rm e}(R)$ is in the polar region (the minimum temperature is here (see Section 3.7), and the high-speed solar wind forms), and the slowest decrease is in the streamer (the highest temperature is at the streamer base, and the low-speed solar wind forms in the region of open field lines in the streamer).

  %{\S}{\bf --- Electron density in radial and nonradial scans of srimer and saddle} \\
Fig. 11(g,h) presents comparison between distributions of electron density along the nonradial streamer and saddle and those along radii from the centre of the streamer base and saddle at $R=1.2R_{\rm o}$. From the figure we notice that difference between two types of $N_{\rm e}(R)$ distribution is not big.

\subsection{Temperature of coronal plasma in different coronal structures}%%%%%%%%%%%%%%
  \label{S-equations}
Electrons and various types of positive ions are known to have different temperatures in the corona, and different methods are used to measure them \cite{Badalyan86,Brandt67}. Representative temperature of coronal plasma in the lower corona may be estimated from observations of the white-light corona, resting on simplifying assumptions:\\
  (1) plasma is spherically symmetric and is at hydrostatic equilibrium;\\
  (2) temperature of all charged plasma components is the same;\\
  (3) plasma temperature is independent of distance \cite{Badalyan86,Brandt67}.\\
Under conditions of hydrostatic equilibrium, electron density varies with distance according to the following law \cite{Badalyan86}:

\begin{equation}\label{Eq-11}
N_{\rm e}(r)=N_{\rm e}(r_1)exp{[-\frac{\mu~ m_{\rm H} GM_{\rm o}}{kT}\cdot (1/r_1-1/r)]}
\end{equation}

\begin{figure}    %%%%%%%%%%%%%%%%%% FIGURE 12
   \centerline{\includegraphics[width=1.0\textwidth,clip=]{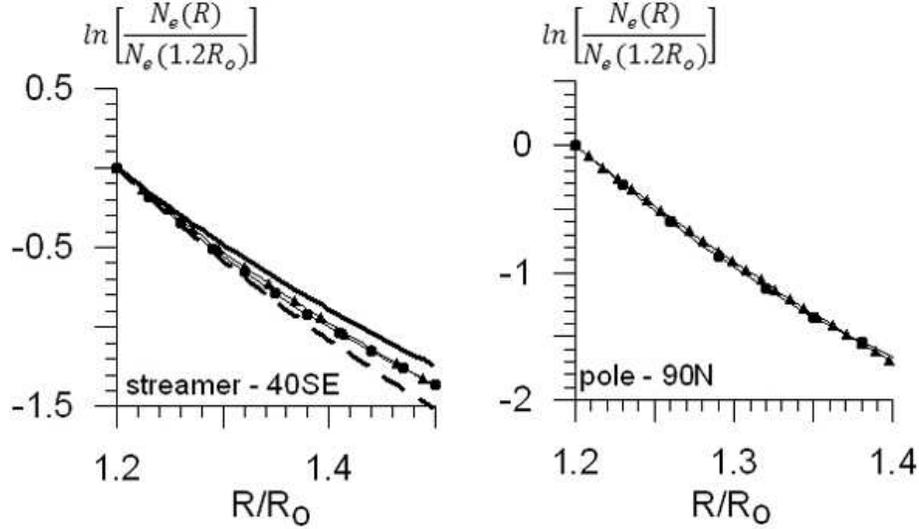}}
   \caption{Determination of coronal plasma temperature at the streamer base and pole ($40SE$ and $90N$, respectively), using matched conditions of radial dependences of electron density (obtained from distributions of the K-corona brightness) and hydrostatic equilibrium in coronal plasma. Line with circles present $N_{\rm e}(R)$ obtained from distribution of the K-corona brightness; lines with triangles - $N_{\rm e}(R)$ obtained from the hydrostatic equilibrium condition ($ln[N_{\rm e}(R/R_{\rm o})/N_{\rm e}(1.2)] = -1.39*10^7/T(1/1.2-1/(R/R_{\rm o}))$). Plasma temperature $\vec{T}$ corresponding to coincidence of two types of distributions, $\approx1.7\cdot10^6К$ at the streamer base $40SE$ and $\approx1.0\cdot10^6K$ at the pole. Solid and dotted lines in the left figure show distributions of $ln[N_{\rm e}(R/R_{\rm o})/N_{\rm e}(1.2)]$ from the hydrostatic equilibrium formula at $T=1.87\cdot 10^6K$ and $1.53\cdot 10^6K$.}
   \label{F-simple}
\end{figure}

Here, $\vec{\mu}\approx0.6$ is the mean molecular weight of particles in the corona, ($\vec{m}_{\rm H}$) is the proton mass, ($\vec{M}_{\rm o}$) is the Sun's mass, ($\vec{T}$) is the plasma temperature, ($\vec{G}$) is the gravitation constant, ($\vec{k}$) is the Boltzmann constant.

  %{\S}{\bf --- Methot of determonation of plasma temperature} \\
We can determine plasma temperature $\vec{T}$ from comparison between the $N_{\rm e}(R)$ distribution from formula (17) and radial dependence of electron density in the given latitude obtained using measurements of the polarisation brightness or K-corona brightness (see Section 3.6.). Plasma temperature $\vec{T}$ (wherein the $N_{\rm e}(R)$ distribution from (17) and the $N_{\rm e}(R)$ dependence obtained from observations of the white-light corona coincide most closely) is taken as plasma temperature in the given latitude. Fig. 12 shows two types of electron density distributions and plasma temperatures, wherein these distributions are closest to each other, for two coronal regions (streamer and pole). From the figure we notice that the maximum plasma temperature ($T\approx1.7\cdot10^6 K$) is at the streamer base $SE$, whereas its minimum temperature ($T\approx10^6 K$) is in the polar region. In other coronal regions, plasma temperature takes intermediate value.

  %{\S}{\bf --- Electron temperature for the eclipse of 1 August 2008} \\
We also measured electron temperature for the eclipse of 1 August 2008 \cite{Habbal10}. According to the results of this work, $T_{\rm e}\approx(1.6\div1.9)10^6~K$ at the streamer base \textbf{SE} and $T_{\rm e}\approx10^6~K$ in polar regions.

  %{\S}{\bf --- changes electron density and electron temperature} \\
Though we use relatively small coronal regions to compare two types of electron density distributions ($R=(1.2\div1.5)R_{\rm o}$), accuracy of the $\vec{T}$ determination is several percent (if the $N_{\rm e}(R)$ distribution obtained from white-light corona data is correct). This is evident from the left-hand figure in Fig. 12 that presents $N_{\rm e}(R)$ distributions for plasma temperature (differing by $\pm10\%$ from the optimal temperature) calculated using formula (14). It is apparent that $N_{\rm e}(R)$ distributions from (17) differ noticeably from the "optimal" as $\vec{T}$ varies within the range $\pm10\%$.

\section{Conclusion} %%%%%%%%%%%%%%%%%%%%%%%%%%%%%%%%%%%%%%%%
      \label{S-Conclusion}
To study the solar corona during eclipses, a new telescope has been constructed: the polarised radiation of the white-light corona passes simultaneously through three polarisers by one objective. Transmission directions of the polarisers are turned through $0^\circ$, $60^\circ$ and $120^\circ$ to the chosen direction. As a result, three images of the white-light corona are obtained. Except for the channels transmitting radiation through polarisers in the telescope, there is one more channel without polariser at the entrance.
This telescope was used to observe the solar corona during the eclipse of 1 August 2008. When making observations, we had two goals: (1) to test the new instrument under the conditions of real observation of the white-light corona; (2) to gain information about coronal features on eclipse day.
The coronal features that we obtained agree with results of coronal observations during other eclipses, as evidenced by testing of our telescope. At the same time, we revealed some disadvantages of the instrument: noticeable lens flares and other things supposed to be eliminated or minimised in the future.

We developed method for primary processing of original coronal images, which implied elimination of the dark current, correction of images for the flat field, overlay of images with one exposition and those with some exposures, etc.

We obtained distributions of the polarisation brightness, K-corona brightness, degree of the K-corona polarisation, total polarisation degree of the white-light corona, total polarisation degree comprising spurious radiation in the plane of the sky. These features were compared in different coronal structures (equator, streamers, saddle, polar regions, polar plumes). Since some coronal structures are nonradial, we compared radial scans of the total polarisation degree and distributions of the total polarisation degree along these structures.

We revealed and studied a coronal structure with special polarisation features ("saddle") located between two streamers. This structure was found out to precede occurrence of an isolated coronal hole 1-2 days later on the eastern limb.

Using distributions of the polarisation brightness and K-corona brightness, we obtained distributions of electron density of coronal plasma, depending on the latitude and radius in the plane of the sky. Features of these $N_e(R,\theta)$ distributions generally agree with those of similar distributions obtained from observations of the white-light corona during other eclipses.

We determined plasma temperature in different coronal structures on the assumption that there is a hydrostatic equilibrium in the lower corona. The maximum temperature of plasma was found out to be at the base of the brightest streamer, whereas its minimum temperature was observed in polar regions.

%%%%%%%%%%%%%%%%%%%%%%%%%%%%%%%%%%%%%%%%%%%%%%%%%%%%%%%%%%%%%%%%%%%%%%%%%%%
\begin{acks}
Authors used data from SOHO/LASCO and SOHO/EIT instruments. SOHO is the project of international cooperation between ESA and NASA. We are grateful to specialists from MLSO for making MarkIV data available. Also we are grateful to WSO for given opportunity to use results of calculations of magnetic field on source surface. The authors thank M.V. Eselevich for the programme for calculating electron density by the inversion method \cite{Hayes01}.
\end{acks}

%%% BIBLIOGRAPHY %%%%%%%%%%%%%%%%%%%%%%%%%%%%%%%%%%%%%%%%%%%%%%%%%%%%%%%%%%%

\end{article}
\end{document}